\documentclass{article}
\usepackage{graphicx}  
\usepackage{amssymb}
\begin{document}
\begin{flushright}
{\bf {DFUB 16/2002 \\ Bologna 12/11/2002 }}
\end{flushright}
\vspace{0.5cm}
\begin{center}
{\Large{\bf {The ANTARES neutrino telescope}}} \par
\vspace{1cm}
Y.~Becherini for the ANTARES collaboration \\ 
Physics Department, University of Bologna and INFN, \\
viale C. Berti Pichat 6/2, \\ 
I-40127 Bologna, Italy \\ 
Yvonne.Becherini@bo.infn.it \par
\vspace{0.5cm}
Proceedings for the International School of Physics ``Enrico Fermi'', \\
CLII course ``Neutrino Physics'', \\
Varenna, Italy, 23 July - 2 August  2002.

\end{center}

\begin{abstract}
The ANTARES collaboration is building a deep underwater neutrino Cerenkov telescope
at 2400 m which will be located off the Mediterranean sea coast near Toulon, France.
The main scientific aims of the experiment are the detection of high energy upgoing muons coming from 
astrophysical neutrinos, indirect dark matter searches, the study of atmospheric neutrino oscillations.
The detector will be completed in the end of 2004.
\end{abstract}
\section{Introduction}
Neutrino telescopes use the detection of upward-going muons as a  signature of $\nu_{\mu}$ CC
interactions in the matter below or inside the detector.  
The ANTARES muon detection medium is sea water through which the muon emits Cherenkov light. 
The light detection allows the determination of the muon trajectory.
The detection technique requires discriminating upward going muons against the much higher flux of 
downward atmospheric muons, and for this reason the detector is installed deep underwater.
Since neutrino cross sections are very small, the detector mass must be very large.

\section{ANTARES scientific programme}
The ANTARES scientific programme can be summarized in three main subjects: neutrino astronomy,
neutralino dark matter searches and atmospheric neutrino oscillation studies. \par
{\bf{Neutrino astronomy.}} Photon astronomy gives an essential contribution to the understanding of the
Universe, it has many advantages because photons are produced in large quantities and can be detected
over a wide energy range, but it also has some disadvantages.
Some hot and dense regions of the sky, like star cores,
are completely opaque to photons; moreover,
high energy photons ($E>10$ TeV) can interact with IR radiation, with
the cosmic microwave background and radio waves through pair production, 
preventing the observation of regions at distances larger than $50$ Mpc.
Many models predict that sources in the Universe may emit high energy ($\gtrsim 1$ GeV) neutrinos.
These neutrinos could be produced by cosmic accelerators, such as gamma ray bursters (GRB),
active galactic nuclei (AGN), supernova remnants, and binary systems.
Neutrino astronomy could open a new exploratory window in the Universe, complementing
high energy gamma ray astronomy. \par
{\bf{Indirect Neutralino dark matter searches.}} Neutralinos could be part of the dark matter halo
in our Galaxy and could appear as Weakly Interacting Massive Particles (WIMP).
Neutralinos could slow down by elastic collisions in celestial bodies 
like the Earth, the Sun or the Galactic
Centre and could gravitationally become trapped in their cores.
Here, neutralino pair annihilation could take place, producing Standard Model particles decaying into
neutrinos.
In a neutrino telescope like ANTARES, such processes could be observed as an excess signal of
induced muons from the celestial body direction. \par
{\bf{Neutrino oscillation studies.}} Evidence for atmospheric neutrino oscillations has been shown by
the MACRO \cite{macro}, SuperKamiokande \cite{sk}, and Soudan 2 \cite{soudan2} collaborations.
ANTARES will have a muon energy threshold of $\sim 5$ GeV,
making the detector capable to measure the atmospheric muon neutrino flux.
An estimate of the ratio $E_\nu/L$, where $E_\nu$ is the neutrino energy and $L$ is the distance
travelled by the neutrino from the production point to the interaction point, will be done to investigate
atmospheric muon neutrino deficit.

\begin{figure}
\begin{center}
\includegraphics[width=8.5cm]{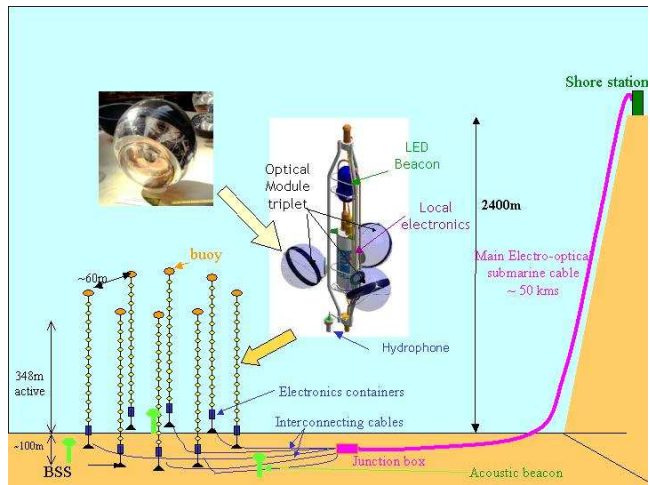}
\end{center}
\caption{\label{fig1} \small Schematic view of the ANTARES underwater neutrino telescope. 
The final detector will consist of a minimum of $10$ strings.}
\end{figure}

\section{Detector design}
The ANTARES neutrino telescope will be installed in the next two years at $2400$ m depth in the 
Mediterrenean sea, $37$ km off
shore of La Seyne sur Mer, near Toulon, France.
Many environmental site measurements have been done to test sea bed, sea currents, water transparency, 
bio-fouling, sedimentation \cite{bio}.
\par
The detector consists of $900$ optical modules in $10$ identical $400$ m long mooring lines (``strings'')
anchored to the sea bed. The distance between the strings will be about $60$ m.
Each string holds $30$ storeys separated by a distance of $12$ m; each storey consists of 3 optical 
modules oriented at $45^\circ$ below the horizontal, see the centre of Fig. 1.
The storeys are interconnected by an electromechanical cable. \par
An ANTARES optical module (OM) \cite{om} is composed of a $17$'' diameter pressure resistant 
glass sphere
containing a $10$'' Hamamatsu photomultiplier tube with its associated electronics.
The angular acceptance of the optical modules is broad, and it falls to half maximum at $70^\circ$
from the axis.
The OM configuration allows to detect light with high efficiency in the lower hemisphere, and has some
acceptance for downgoing muon tracks.
The relative positions of all optical modules in the detector are given in real time by an acoustic
positioning system and by compasses and tilt-meters installed along the line which allow the 
reconstruction of the shape of the line and the orientation of each storey. \par
The electronics related to a given storey is contained in a local control module (LCM), 
which contains the boards for readout, DAQ, power, clock and trigger.
At the base of each string there is a string control module (SCM), which contains the electronics 
concerning the Slow Control, clock, and instruments for acoustic positioning and measurements of sea 
properties. The individual SCMs are linked to a common junction box (JB) by electro-optical cables which are
connected using a submarine.
A deep sea telecommunication cable links the JB with a shore station where the data
are filtered and written to disk. \par
A prototype of the detector string ({\it {sector line}}) will be deployed in
November 2002. 

\begin{figure}
\begin{center}
\mbox{
\hspace{-1.0cm}
\includegraphics[width=7.5cm]{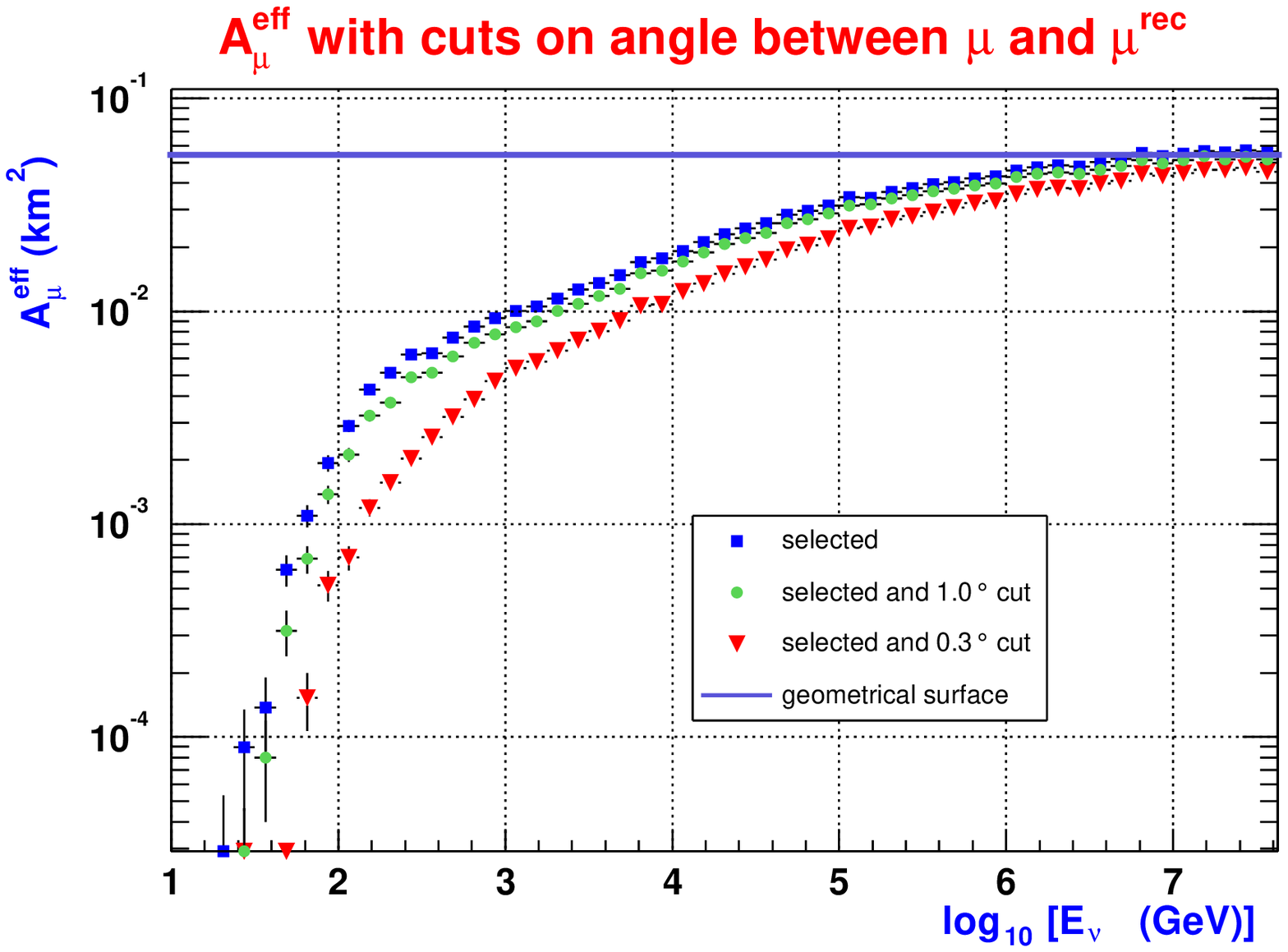}
\hspace{-1.0cm}
\includegraphics[width=7.5cm]{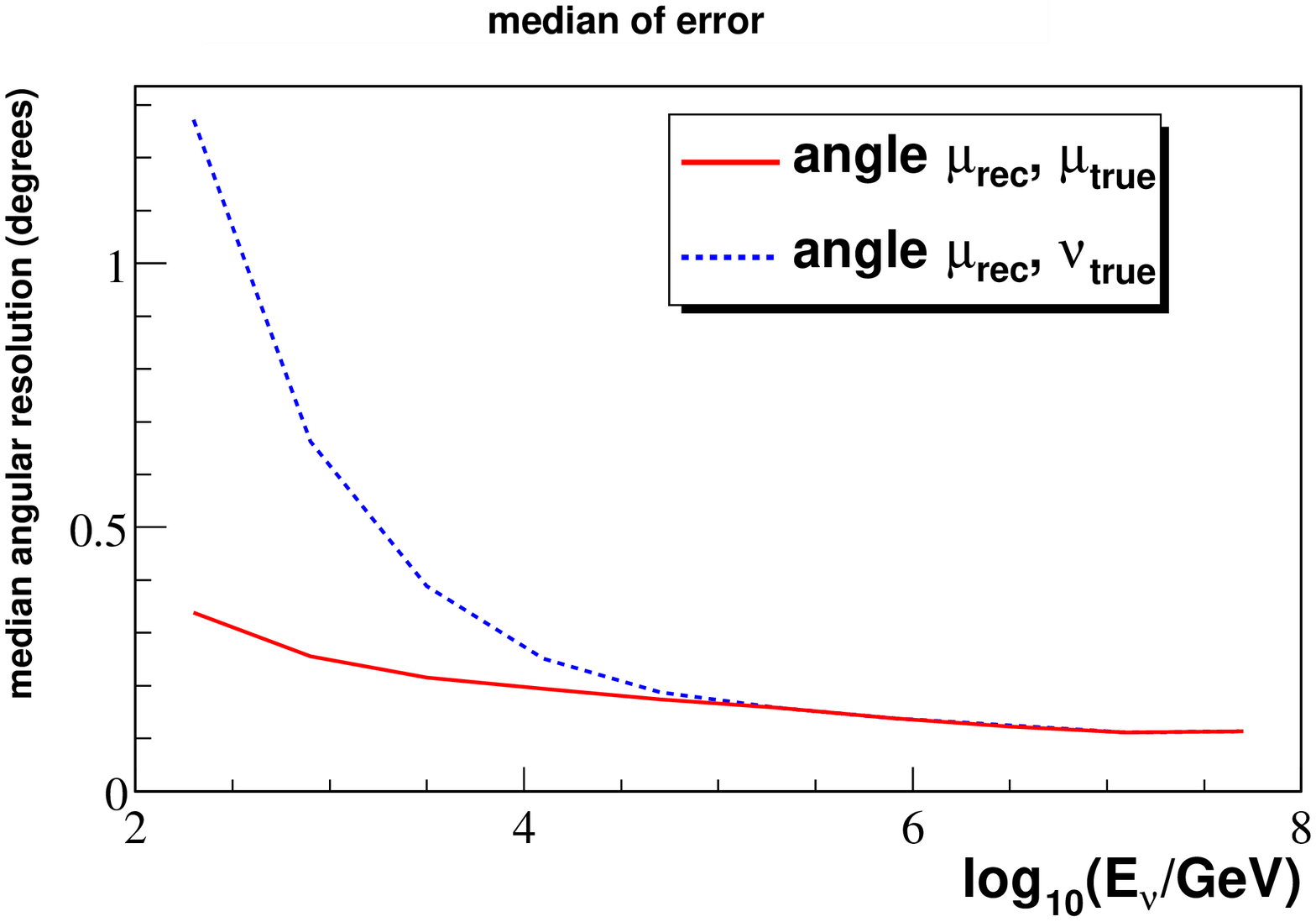}}
\end{center}
\caption{\label{fig2} \small {\bf {Left:}} effective area as a function of neutrino 
energy after applying quality cuts, dots are for all selected events, squares and triangles 
include the requirement
that the reconstruction error is lower than $1^\circ$ and $0.3^\circ$, respectively. 
The solid line represents the geometrical area.
{\bf{Right:}} median value of the distribution of the angle between the reconstructed
muon and the generated muon (solid line), and between the reconstructed muon and the parent neutrino 
(dashed line) versus the neutrino energy \cite{montaruli}.}
\end{figure}


\section{Detector expected performances}
The relevant parameters which characterise a neutrino telescope are its effective area, total mass,  
its angular resolution (for astronomy) and energy resolution. 

{\bf {Effective area.}} 
Fig. 2 on the left shows the effective area computed using a Monte Carlo simulation of isotropic 
neutrino events as a function of neutrino energy 
after quality cuts on the reconstruction and how the effective area depends on the required
pointing accuracy.
For a typical E$^{-2}$ cosmic neutrino spectrum
$96\%$ ($72\%$) of the events are reconstructed
with an error smaller than $1^\circ$ ($0.3^\circ$). 

{\bf {Angular resolution.}} The intrinsic angular resolution of the telescope is 
defined as the median angular separation between the real and the reconstructed muon track.
The angular resolution of a neutrino telescope depends on reconstruction algorithms, 
selection programs and timing accuracy.
In Fig. 2 (on the right) the median value of the distribution
of the angle between the reconstructed
muon and the simulated muon, and between the reconstructed muon and the parent neutrino 
versus the neutrino energy are shown.
Below 10 TeV the median angle between the muon and the neutrino is dominated
by the kinematics of the interaction, while at larger energies it is limited
by the intrinsic angular resolution. 
At higher energies the neutrino pointing accuracy $<0.2^\circ$. 
This estimate takes into account of the light scattering in water.

\begin{figure}
\begin{center}
\includegraphics[height=6cm]{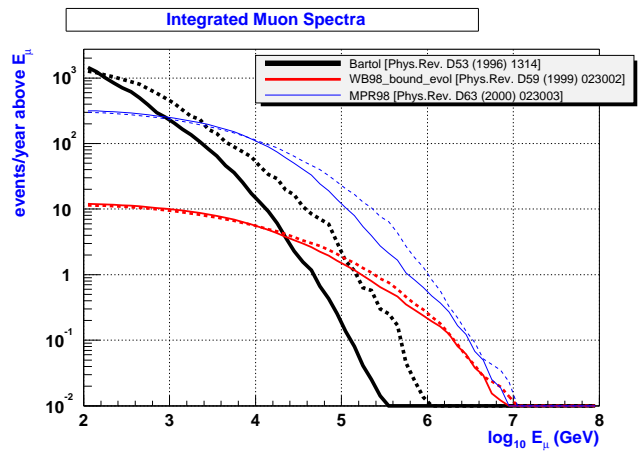}
\end{center}
\caption{\label{fig3} \small Comparison between the predicted (solid lines) and reconstructed (dashed lines) 
integrated spectra for the atmospheric neutrino flux, and for two different astrophysical 
neutrino fluxes WB98 and MPR98.}
\end{figure}

{\bf {Energy resolution.}}  
Below 100 GeV, the muon energy is determined from the range in the detector, 
while above $100$ GeV the energy is estimated by the quantity of light detected
by the optical modules.
For $E_\mu=1$ TeV the muon energies are reconstructed within a factor of $4$, decreasing to a factor of 
$3$ at $10$ TeV, reaching the value of $2$ for $10<E<10^7$ TeV.
Fig. \ref{fig3} shows the comparison between the generated and the reconstructed integrated muon spectra 
for several neutrino flux models. 

\begin{thebibliography}{0}
\bibitem{macro} S. P. Ahlen et al., Phys. Lett. B 357 (1995) 481. 
P. Ambrosio et al., Phys. Lett. B 434 (1998) 451; hep-ex/0206027. 
\bibitem{sk} SK Coll., Y. Fukuda et al., Phys. Rev. Lett. 81 (1998) 1562; 85 (2000) 3999.
\bibitem{soudan2} Soudan 2 Coll., W. W. M. Allison et al., Phys. Lett. B 449 (1999) 137.
\bibitem{bio} P. Amram et al., Astr. Phys. 13 127-136 (2000); Amram P. et al., astro-ph/0206454.
\bibitem{om} P. Amram et al., NIM A 484 (2002) 369.
\bibitem{montaruli} T. Montaruli, ANTARES collaboration, astro-ph/0207531.


\end{thebibliography}
\end{document}